# Conformational catalysis of cataract-associated aggregation by interacting intermediates in a human eye lens crystallin

Eugene Serebryany, Rostam M. Razban, and Eugene I. Shakhnovich

Department of Chemistry and Chemical Biology, Harvard University, Cambridge MA 02138 USA

**Abstract**

Human γD-crystallin (HγD) is an abundant two-domain protein concentrated in the core region of the eye lens. Destabilizing mutations and post-translational modifications in the N-terminal domain of γ-crystallins are linked to onset of aggregation that causes cataract disease (lens turbidity). WT HγD greatly accelerates aggregation of the cataract-related W42Q variant, without itself aggregating. The mechanism of this "inverse prion" catalysis of aggregation remained unknown. Here we provide evidence that it proceeds via an early unfolding intermediate with an opened domain interface, which enables transient dimerization of WT and mutant, or mutant and mutant, HγD molecules at their C-terminal domains. This dimerization deprives the N-terminal domain of intramolecular stabilization by the domain interface and thus promotes its conversion to a distinct, aggregation-prone partially unfolded intermediate. This mechanism can be generalized to explain how surprising reactions, such as conformational catalysis of misfolding, may arise from simple domain-domain interactions in multidomain proteins.

**Significance**

Most known proteins in nature consist of multiple domains. Interactions between domains may lead to unexpected folding and misfolding phenomena. This study of human γD-crystallin, a two-domain protein in the eye lens, revealed one such surprise: conformational catalysis of misfolding via intermolecular domain interface "stealing." An intermolecular interface between the more stable domains outcompetes the native intramolecular domain interface. Loss of the native interface in turn promotes misfolding and subsequent aggregation, especially in cataract-related γD-crystallin variants. This phenomenon is likely a contributing factor in the development of cataract disease, the leading worldwide cause of blindness. However, interface stealing likely occurs in many proteins composed of two or more interacting domains.

**Introduction**

Beyond the native and fully unfolded states, most proteins can adopt partially unfolded or misfolded intermediates. Intermediates are especially important for multidomain proteins, which comprise most eukaryotic proteins yet only a small minority of biophysical studies to-date (1-3).

These intermediates are often rare or transient and thus challenging to investigate, yet they have large effects on the protein's stability, evolution, and function. Misfolded intermediates can dominate folding pathways (4-7), drive evolution of the coding sequence (8-10), and cause proteins to assemble into aggregates linked to common diseases (11-17). Post-translational modifications often affect the relative stabilities of intermediates within the conformational ensemble (18-24). However, where a protein is present at a high concentration, the question arises: Can distinct intermediate states interact or affect one another in a functionally significant way? Here we present experimental evidence that domain interface opening in one molecule of a stable two-domain protein can catalyze domain misfolding in another.

Lens crystallins are extremely long-lived. Human γD-crystallin (HγD) in the core region of the eye lens is synthesized *in utero* and never replaced thereafter. It is extremely thermodynamically and kinetically stable, likely an evolutionary adaptation to resist aggregation (17, 25-28). However, crystallins accumulate damage throughout life, even as the cytoplasm of aged lens cells becomes increasingly oxidizing (29-36). Eventually, aggregation-prone partially unfolded intermediates arise, notably those locked by non-native disulfides (33, 37-43). Cataract is lens opacity that results when these misfolded crystallins assemble into light-scattering aggregates, making the lens turbid. Cataract is a disease of aging, affecting 17% of all people aged 40 and over (44, 45), though many congenital or early-onset mutations are also known (46, 47). Due to the high cost of surgery and lack of a therapeutic treatment, cataract remains the leading cause of blindness in the world (45, 48).

HγD consists of two homologous double-Greek key domains joined by a non-covalent domain interface (49, 50). This interface contributes significantly to the stability of the N-terminal domain (50-52). Destabilization and misfolding of the N-terminal domain under oxidizing conditions leads to an aggregation-prone intermediate state kinetically trapped by a non-native internal disulfide bond (41). Surprisingly, the highly stable wild-type HγD promotes aggregation of its destabilized W42Q variant without itself aggregating – an interaction that may permit even a minor fraction of damaged molecules to cause significant lens turbidity (53).

We have previously reported the crucial interplay between misfolding and disulfide exchange in the aggregation of cataract-associated HγD variants (42). Specifically,. we showed that HγD molecules can exchange disulfide bonds dynamically in solution, which leads to the "hot potato" model: if a molecule with a destabilized N-terminal domain accepts a disulfide bond from a more stable molecule, aggregation of the destabilized variant will ensue (42). However, we now demonstrate that the wild-type protein also promotes aggregation of multiple cataract-related variants of HγD by a distinct mechanism independent of "hot potato" disulfide exchange. This novel mechanism is associated with a partially unfolded intermediate in the wild-type protein wherein the domain interface is opened. Population of this intermediate enables a transient "interface stealing" interaction between molecules such that the N-terminal domain of the aggregation-prone variant loses its interface with the C-terminal domain and hence the kinetic and thermodynamic stabilization derived from that interface. Thus, the interaction facilitates unfolding of the mutant's N-terminal domain leading to its conformational transformation and aggregation. We refer to this new class of protein-protein interactions as conformational catalysis of misfolding via transient interaction of non-native states.

Results

**Conformational catalysis of W42Q aggregation**

The W42Q variant of HγD mimics UV-induced damage to a Trp residue and shows significant propensity to aggregate at physiological temperature and pH, similar to the congenital-cataract mutant W42R (41, 43, 54). This aggregation process requires formation of a non-native internal disulfide bond (Cys32-Cys41), which traps the protein in a non-native "hairpin extruded" conformation (41). Because of this requirement of oxidation combined with misfolding, aggregation does not occur in the absence of oxidizing agent (41, 42).

When the oxidizing agent glutathione disulfide (GSSG) was present in the buffer, aggregation of W42Q proceeded readily at 37 °C and pH 7 in a concentration-dependent manner (**Fig. 1A**). We have previously shown that a disulfide in wild-type HγD (Cys108-Cys110) is able to replace GSSG as oxidizing agent for W42Q aggregation (42). We investigated whether presence of WT produced any further effect on W42Q aggregation when excess GSSG was also present in the buffer. Presence of WT led to dramatic, [WT]-dependent acceleration of W42Q aggregation despite the excess GSSG (**Fig. 1B**). Adding more GSSG did not significantly affect aggregation (**Fig. 1C**). This finding indicated that there exists a distinct, redox-independent mechanism by which WT promotes aggregation of W42Q which is not related to disulfide transfer from WT to aggregation prone variant.

We previously found that, in the absence of any other oxidant in the buffer, removal of Cys108 or Cys110 from the WT rendered it unable to promote aggregation of W42Q (42). However, when excess GSSG was present in the buffer, removing any Cys residue from the WT protein by site-directed mutagenesis did not abolish WT-induced acceleration of the aggregation rate, and often actually enhanced it (**Fig. 1D**). Furthermore, it did not matter whether the WT protein was oxidized or reduced – as long as a source of disulfides was present in excess, aggregation in the WT/W42Q mixture was rapid (**Fig. S1**). Thus, redox chemistry was not the basis for the accelerated aggregation.

To investigate whether the increase in solution turbidity was due to increased W42Q aggregation or to co-aggregation of WT and W42Q, we determined the composition of the aggregated fraction. To that end we used a previously established PEGylation assay (42) to determine the total number of Cys residues per protein molecule in soluble vs. aggregated fractions of W42Q mixed with one of two multi-Cys mutants in the WT background. The variants used were C18T/C41A/C78A ("CCC") and C18T/C78A/C108S/C110S ("CCCC"). Like WT, both variants promoted W42Q aggregation (**Fig. 1D**). Importantly, both CCC and CCCC variants remained entirely within the soluble supernatant fraction at the end of the aggregation reaction, while the majority of W42Q transferred to the aggregated fraction (**Fig. 1E**). All samples were reduced prior to PEGylation, hence W42Q reacted with 6 molecules of PEG-maleimide per protein molecule; a minor band at 4 PEG-maleimide per protein molecule indicates reduction was not complete. As a complementary experimental approach, we examined the composition of soluble and aggregated fractions in W42Q/WT mixtures by MALDI-TOF mass spectrometry, which likewise showed only W42Q in the aggregated fraction and WT in the supernatant (**Fig. S2**). We conclude that the increase in solution turbidity in W42Q/WT or W42Q/CCC or W42Q/CCCC mixtures was not due to co-aggregation, but rather to increased aggregation of the W42Q variant.

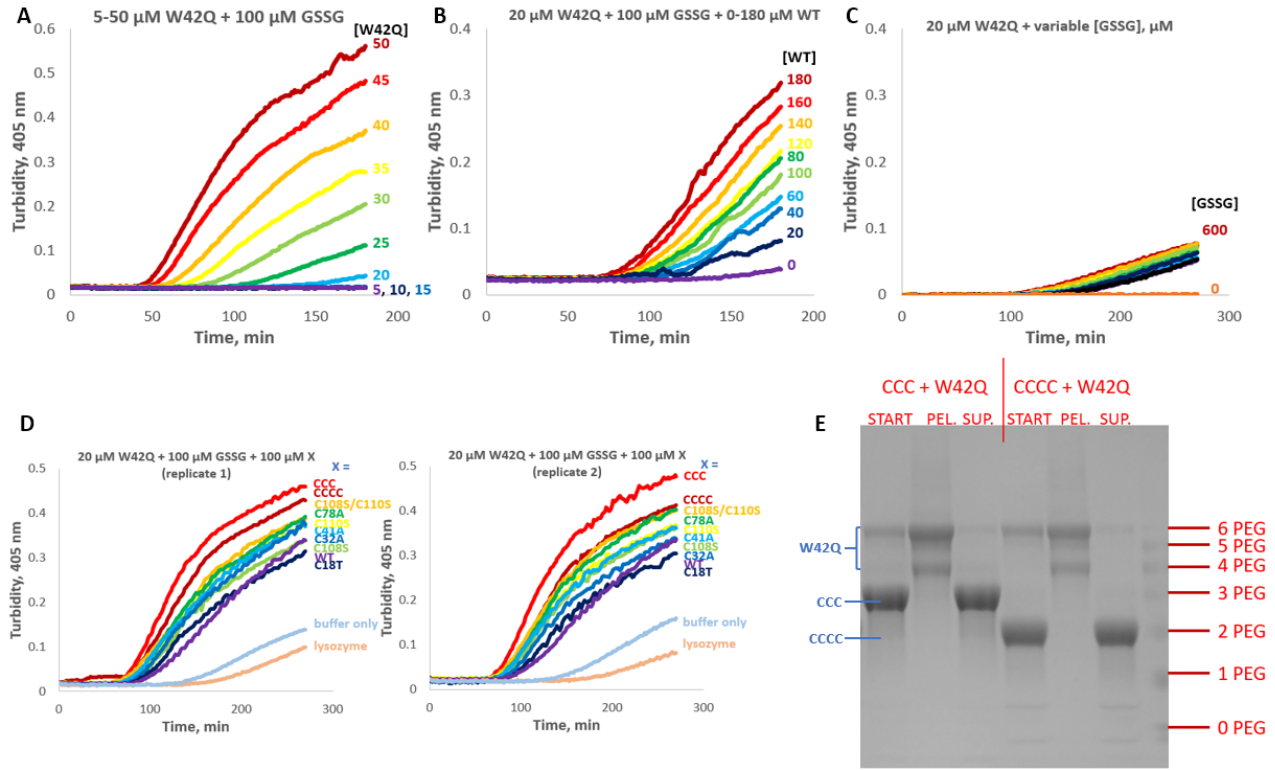

**Fig. 1: WT HγD promotes aggregation of the W42Q variant in a catalytic, redox-independent manner.** (A) W42Q HγD aggregated in a concentration-dependent manner in the presence of excess glutathione disulfide (GSSG). (B) WT promoted aggregation of W42Q in a [WT]-dependent manner. (C) Further excess of GSSG did not significantly affect W42Q aggregation. (D) Mutagenesis of any given Cys residue, or the combination of C18T/C41A/C78A ("CCC") or C18T/C78A/C108S/C110S ("CCCC") in the WT background did not abolish its ability to promote W42Q aggregation at saturating GSSG, and indeed enhanced it. A control protein (lysozyme) did not promote W42Q aggregation. Two replicates are shown as illustration of reproducibility. (E) Cys-counting analysis of the starting ("START"), aggregated ("PEL"), and supernatant ("SUP") fractions of W42Q/CCC and W42Q/CCCC mixtures. The W42Q variant contains 6 Cys residues and thus reacts with a maximum of 6 PEG-maleimide equivalents (4 if the internal disulfide required for aggregation has not been reduced). The CCC and CCCC variants contain only 3 and 2 Cys residues, respectively. Neither CCC not CCCC was detectable in the aggregated fraction, which is composed entirely of W42Q, indicating that the aggregation-promoting activity was catalytic.

We hypothesized that presence of WT accelerated conformational dynamics of W42Q making it more amenable to convert to an aggregation competent conformation containing the non-native disulfide bond 32-SS-41 (41). To that end, time-resolved PEGylation was used to probe the effect of conformational catalysis on the structure and conformational dynamics of the W42Q N-terminal domain. Since the N-terminal domain contains four Cys residues, three of which are buried in the domain core and the fourth in the domain interface, Cys accessibility is a good probe for foldedness of the domain. Samples from W42Q and a W42Q/CCCC mixture incubated in the presence of PEG-maleimide at 37 °C in native buffer were taken at fixed intervals, the PEGylation reagent quenched with free cysteine, and the proteins resolved by SDS-PAGE to quantify the distribution of singly and multiply PEGylated molecules (**Fig. 2** and **Fig. S3**). The CCCC sample contains only two Cys residues, one buried in the N-terminal core and the other in the domain interface. In control samples of CCCC alone only one Cys residue was found to be reactive during an hour-long incubation (**Fig. 2**, *gray*). We assign this to the residue buried in the domain interface, Cys41; the kinetics of its reactivity is related to the kinetics of interface opening in this variant. By contrast, the W42Q protein reacted extensively with PEG-maleimide during the course of the incubation, and both initial and full PEGylation was achieved more rapidly in the W42Q/CCCC mixture than in W42Q alone (**Fig. 2**). The first reactive site (+1 PEG) is assigned to Cys110, which is the only natively exposed and the most reactive thiol (35, 42, 49). The second reactive site in W42Q (+2 PEG) may be Cys41, since its solvent exposure requires only opening of the domain interface, which occurs relatively easily (41, 52, 55). However, no definitive assignment has been made. Kinetics of reactivity at the second site, as well as for full unfolding (+6 PEG) were significantly faster in the presence vs. absence of the conformational catalyst. Flux through the intervening numbers of exposed thiols was comparable, however, suggesting that following exposure of the first (or possibly second) natively buried Cys residue unfolding becomes cooperative. We conclude that the aggregation-catalyzing interaction between W42Q and CCCC has a structural basis: it increases the unfolding rate of the W42Q N-terminal domain.

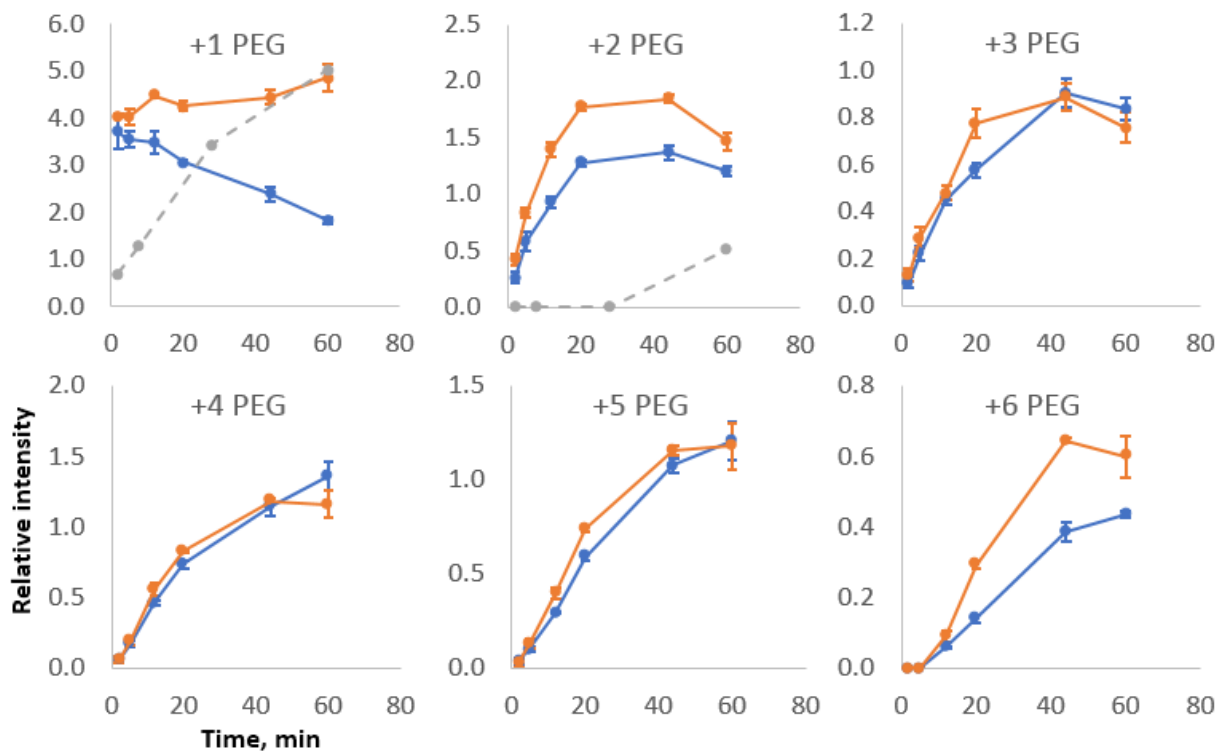

**Fig. 2: WT protein accelerates transient unfolding of W42Q.** W42Q unfolded more rapidly when CCCC was present (*orange*) than without CCCC (*blue*). PEGylation is a measure of solvent accessibility of buried Cys residues. Error bars show S.E.M. of 3-4 replicate measurements with independent quantitation by SDS-PAGE. Control experiments with CCCC alone show the reactivity of its only two Cys residues (*gray*). PEGylation gels are shown in **Fig. S3**.

**Kinetic model of conformationally catalyzed HγD aggregation**

Our findings point to a peculiar mechanism whereby WT protein can catalyze aggregation of mutant form without aggregating itself. To explore this effect more quantitatively, we developed a minimalist kinetic model as described below.

For an aggregation-prone mutant (***MUT***), we consider the following four conformational states. ***N*** is the native state of the protein, which we define here as any state that is neither aggregation prone nor likely to catalyze aggregation. $I_{CAT}$ is an early conformational intermediate that is capable of catalyzing aggregation but is not itself aggregation prone. $I_{SS}$ is the disulfide-locked partially unfolded intermediate that is the aggregation precursor. ***A*** is the high molecular weight aggregate. The conformational conversions for conformationally catalyzed aggregation follow this scheme:

$$N \underset{k_{-CAT}}{\overset{k_{CAT}}{\rightleftarrows}} I_{CAT} \xrightarrow{k_{SS}} I_{SS} \xrightarrow{k_A} A$$

Only *MUT* proceeds from $I_{CAT}$ to $I_{SS}$. Since prior experiments have shown that $I_{SS}$ in W42Q forms at temperatures where no unfolding is observed (53), requires a non-native disulfide (41), and partitions quantitatively to the aggregated state (42), we assume that the rate-limiting step for aggregation is the $I_{CAT}$ to $I_{SS}$ conversion. We have previously shown that solution turbidity varies monotonically with amount of aggregated protein (53), so the rate of turbidity rise at a given early aggregation time point is proportional to the rate of aggregation. Equilibration between *N* and $I_{CAT}$ is rapid relative to the $I_{CAT}$ to $I_{SS}$ conversion, so we define the equilibrium constant $K_{CAT} = k_{CAT}/k_{-CAT}$. Note that the value of each rate constant above is specific to the given mutant. We now consider the simplest kinetic models that describe the following three cases:

**Case I: No catalysis**

The simplest model in this case involves direct conversion of *N* to $I_{SS}$ without the need for the catalytic intermediate.

$$\frac{d[I_{SS}]}{dt} = k_{SS}[N] - k_A[I_{SS}]^n$$

$$\frac{d[A]}{dt} = k_A[I_{SS}]^n$$

The maximum rate of aggregation occurs when *[$I_{SS}$]* is highest, that is, *d([$I_{SS}$])/dt = 0*, so the dependence of maximum aggregation rate on starting concentration of the mutant is linear:

$$\left.\frac{d[A]}{dt}\right|_{max} = k_{SS}[N]$$

**Case II: Self-catalysis**

In this case the rate-limiting step is facilitated by an interaction between mutant molecules in the catalytic intermediate state. Hence:

$$[I_{CAT}] = K_{CAT}[N]$$

$$\frac{d[I_{SS}]}{dt} = k_{SS}[I_{CAT}]^m - k_A[I_{SS}]^n$$

$$\frac{d[A]}{dt} = k_A[I_{SS}]^n$$

$$\left.\frac{d[A]}{dt}\right|_{max} = k_{SS}[I_{CAT}]^m \propto [N]^m$$

Since the maximum aggregation rate occurs very early in the kinetic trace, *[N]* is approximately equal to the starting concentration. The $K_{CAT}$ is variant-specific. The simplest model is that the catalytic complex consists of only two $I_{CAT}$ molecules, so *m = 2* – the dependence of maximum aggregation rate on starting concentration is quadratic. This is in very good agreement with experimental data shown in **Fig. 3A**. **Fig. S4** shows the full aggregation traces for these variants and demonstrates that a linear fit fails.

**Case III: Heterogeneous catalysis**

This case differs from Case II in only one key respect: Catalysis can be carried out not only by the $I_{CAT}$ state of **MUT**, but also by the equivalent conformation of WT or other non-aggregation prone variant. Using the simple model of a bimolecular catalytic complex, which was validated in Case II, we have:

$$[I_{CAT}] = K_{CAT}[N]$$

$$\frac{d[I_{SS}]}{dt} = k_{SS}([I_{CAT}] + [I_{CAT}^{WT}])[I_{CAT}] - k_A[I_{SS}]^n$$

$$\frac{d[A]}{dt} = k_A[I_{SS}]^n$$

Note that, since the non-aggregation prone variant (which we label "WT" for simplicity) does not proceed to the $I_{SS}$ conformational state, its population of $I_{CAT}$ is determined by an equilibrium process. The value of the equilibrium constant is variant-specific.

$$[I_{CAT}^{WT}] = K_{CAT}^{WT}[N^{WT}]$$

When the concentration of the mutant protein is kept constant, self-catalysis makes a constant contribution to the aggregation rate, while heterogeneous catalysis depends on [WT]:

$$\left.\frac{d[A]}{dt}\right|_{max} = [I_{CAT}](k_{SS}[I_{CAT}^{WT}] + [I_{CAT}]) \propto [N]([N^{WT}] + const.)$$

In this case, the maximum aggregation rate depends linearly on the concentration of the WT protein, which is in excellent agreement with the data in **Fig. 3B**. The constant accounts for the fact that self-catalysis makes a modest but non-negligible contribution to the aggregation rate under these conditions. Note that the value of each rate constant above is specific to the given mutant.

To assess the generality of the findings beyond W42Q, we tested two other variants of HγD: V75D and L5S. Even though the three mutations are in distal parts of the N-terminal domain's sequence, all three are associated with cataract (41, 46, 56-58), and all three replace core hydrophobic residues with hydrophilic ones, thus destabilizing the N-terminal domain (see **Table 1**). As expected, each variant showed redox-dependent aggregation comparable to that of W42Q, although higher protein concentrations were necessary to achieve comparable aggregation rates (**Fig. 3A**). This variation can be rationalized structurally, as described below.

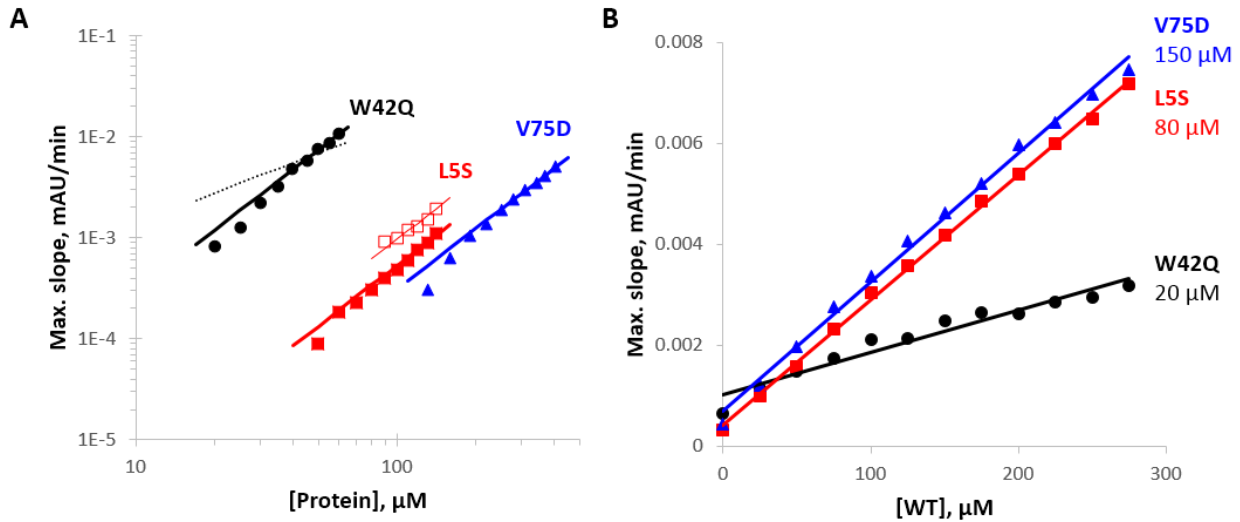

**Fig. 3. Dependence of maximal slope of kinetic turbidity curves on initial concentrations** (in µM) for homogeneous aggregation of aggregation-prone W42Q, V75D, and L5S variants (self-catalysis, A) and (B) heterogeneous catalysis of the same aggregation-prone variants by added WT. The concentration-dependence graphs in (A) are fit to lines of slope = 2 on the log-log plot, corresponding to quadratic parabolas in linear scale. The dashed line represents best linear fit (slope = 1) for comparison. Since a gentle acceleration of aggregation rate was observed at later times in the case of L5S, open squares show rates for this second phase. In (B) fits are linear as predicted by theory. WT appeared to be a stronger catalyst of L5S and V75 aggregation, but the distinction could be due to higher concentrations of these variants (noted on the plot) than for W42Q. (See Discussion.)

**Structural model of catalysis of HγD aggregation**

Examination of the structure, as well as previous simulations (41), indicated that the only partially unfolded conformation expected to be accessible in WT HγD under physiological conditions is the one where the cores of both domains are intact, but the domain interface is opened. *We hypothesize that this conformation with open domain interface is active in conformational catalysis.* The free energy of the domain interface has been estimated at just 4 kcal/mol (52), so WT is expected to populate this intermediate, albeit rarely and transiently, under physiological conditions. The low population of this intermediate (low $K_{CAT}^{WT}$) explains the relatively high concentrations of WT required for catalysis of aggregation (**Fig. 3B**): it is not the native state WT that is catalytically active, but only the minor fraction that adopts the open-interface intermediate conformation at any given time.

To test the hypothesis of conformational catalysis via the interface, we prepared a series of four constructs expected to have distinct interface-opening propensities and evaluated their ability to catalyze W42Q aggregation. Construct 1 was WT bearing an N-terminal KHHHHHHQ tag sequence; this tag has been shown to interact with the domain linker and part of the C-terminal domain (59), so we reasoned it was likely to decrease the population of the open-interface intermediate. Construct 2 was the untagged WT protein as used elsewhere in this study. Construct 3 was the H83P variant, which introduces an extra Pro residue into the domain linker and is expected to increase linker strain. Construct 4 was the double-deletion variant Δ85,86, in which two of the five residues that form the domain linker are deleted, which is expected to result in large linker strain. Indeed, purification of the Δ85,86 variant revealed a modest shift toward earlier elution position by size-exclusion chromatography, consistent with a more open structure, and solutions of the Δ85,86 became cloudy if kept on ice, in contrast to all other crystalline constructs in this study; the cloudiness dissipated completely upon warming the sample to 37 °C or room temperature. The reversible nature of this aggregation process suggests reversible, likely domain-swapped, polymerization exclusively at cold temperatures.

As shown in **Fig. 4A**, the oligomeric state of the four constructs varied in accordance with their expected interface-opening propensities: no detectable dimer in HisWT; an extremely weak dimer peak in WT; a strong one for H83P; and the Δ85,86 variant was almost entirely dimeric. Static light scattering experiments in batch mode confirmed the shift to dimeric structures, as indicated.

The magnitude of conformational catalysis by these four constructs likewise ranked in accordance with their predicted propensity to populate the open-interface intermediate. HisWT was the least active, followed by WT, H83P, and finally Δ85,86 (**Fig. 4B**). Moreover, a mixture of isolated N- and C-terminal domains had a comparable catalytic activity to the Δ85,86 variant (**Fig. 4C**). Thus, the WT protein's catalytic activity was indeed likely related to interface opening.

While both isolated domains produced enhanced turbidity when mixed with the W42Q variant, the reasons for the enhancement were distinct. As shown in **Fig. 4D**, the isolated N-terminal domain co-aggregated with the W42Q variant, whereas the C-terminal domain did not. Likewise, the aggregated fraction in the W42Q/Δ85,86 mixture contained ~10% Δ85,86, a minor but significant amount (**Fig. S5**). It is known that the domain interface in the native HγD structure disproportionately stabilizes the N-terminal domain, both thermodynamically and kinetically (reducing its rate of unfolding) (52, 60). Disrupting the stabilizing domain interface thus likely facilitates misfolding and disulfide-trapping of the N-terminal domain even in the absence of any mutations in the core, leading to the possibility of co-aggregation. Indeed, the isolated N-terminal domain aggregated by itself at slightly elevated temperatures (42 °C), as shown in **Fig. 4E**.

The observation of co-aggregation between the N-terminal domain and W42Q, but no co-aggregation with the C-terminal domain suggests the C-terminal side of the domain interface is responsible for conformational catalysis, and therefore no such catalysis would exist for the isolated N-terminal domain. Our kinetic model predicts linear concentration dependence of maximal aggregation rate in that case, and indeed that is what we observed (**Fig. 4D**), in contrast to the quadratic concentration dependence of self-aggregation in full-length variants (**Fig. 3A**). These data are consistent with the model whereby C-terminal domain interface of W42Q serves as a catalyst for aggregation, facilitating misfolding of the N-terminal domain in a bimolecular interaction. Indeed, molecular dynamics simulations have shown that the W42R mutation destabilizes not only the core of the N-terminal domain, but also the domain interface (55), thus facilitating population of the open-interface conformer and exposure of the C-terminal side of the domain interface. This accounts for the rapid homogeneous aggregation of W42Q compared to L5S (**Fig. 3A**).

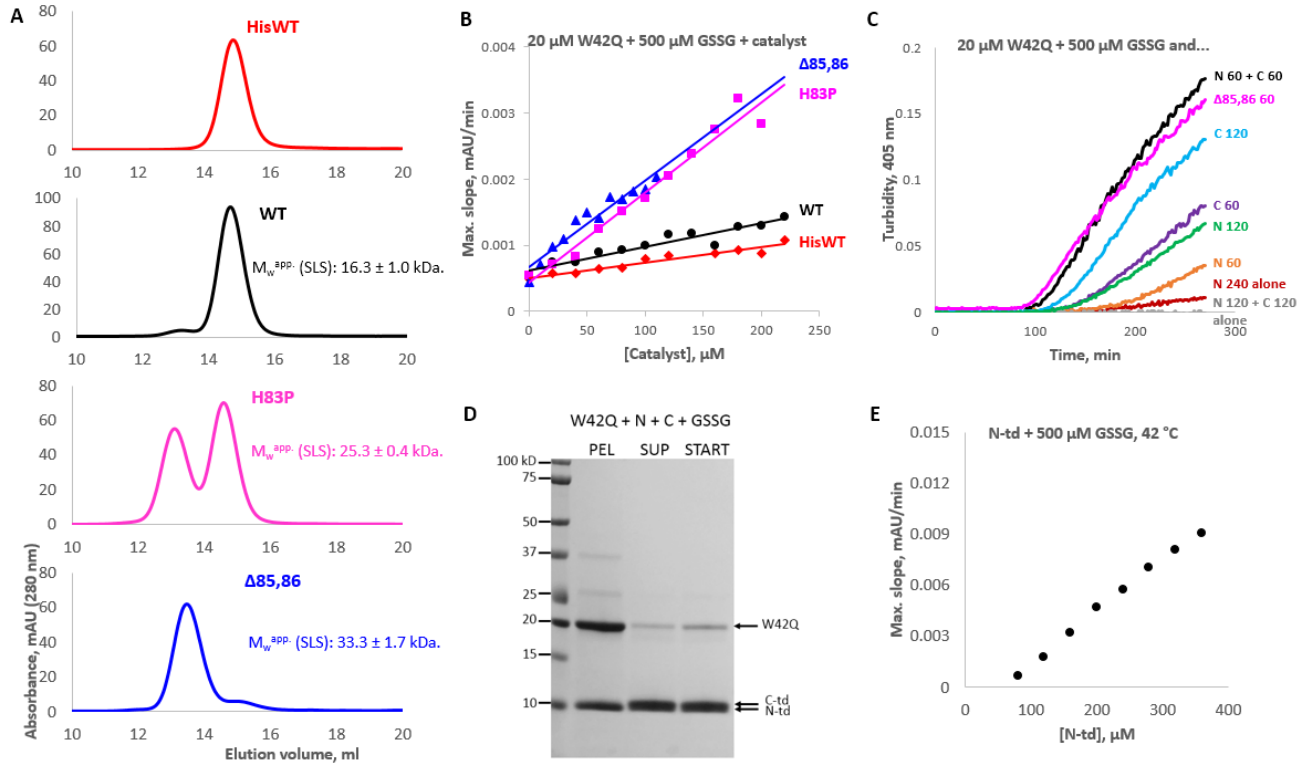

**Fig. 4: Conformational catalysis is associated with the C-terminal side of the domain interface.** (A) Extent of dimerization by size-exclusion chromatography was in the order of predicted interface-opening propensity. Samples were incubated with reducing agent prior to SEC. Dimerization was confirmed by static light scattering, as indicated by the apparent molecular weights in batch mode. (B) Aggregation-promoting ability of His-WT, WT, H83P, and Δ85,86 had the same rank order as expected accessibility of the domain interface. (C) The C-terminal domain promoted aggregation most efficiently, although when N-td and W42Q were both present, aggregation was enhanced.to the same level as the Δ85,86 variant. (D) The N-terminal domain co-aggregated with the W42Q variant in approximately stoichiometric amounts, whereas the C-terminal domain showed no co-aggregation. (E) The N-terminal domain aggregated by itself at a slightly elevated temperature (42 °C), but the concentration dependence of aggregation rate was linear, indicating absence of a catalytic process.

**Table 1: Thermostability of select HγD variants by differential scanning calorimetry**

| Protein | $T_m$ 1 (N-td) | $T_m$ 2 (C-td) |
|---|---|---|
| W42Q [a] | 54.6 ± 0.3 | 71.1 ± 1.7 |
| L5S | 58.3 ± 0.02 | 70.5 ± 0.1 |
| V75D | 60.0 ± 0.01 | 71.0 ± 0.02 |
| **WT** | 81.2 ± 0.1 | 84.1 ± 0.03 |
| **H83P** | 81.5 ± 0.1 | 84.2 ± 0.03 |
| **Δ85,86** | 71.6 ± 0.2 | 74.3 ± 0.04 |

Transition midpoints were obtained by fitting the differential scanning calorimetry traces to a double-Gaussian model using the NanoDSC software. Values are reported as mean ± standard deviation of 2-3 technical replicates. [a]Data from reference (41).

To determine whether the catalytic effect on W42Q aggregation was due to specific features of the domain interface or to its overall hydrophobicity, we studied aggregation of W42Q in the presence of varying concentrations of bovine serum albumin (BSA). BSA natively has ~28 times more hydrophobic surface area per molecule than either of the HγD single domains (and ~60 times more than native HγD WT), as summarized in **Table S6**. Some acceleration of W42Q aggregation was indeed observed in the presence of BSA (**Fig. S7**), but the HγD constructs, despite their much lower overall hydrophobicity, produced much more rapid aggregation than BSA, indicating that the conformational catalysis cannot be explained by mere hydrophobicity.

Investigation of the isolated C-terminal domain provided a structural explanation for this catalytic mode of action of the C-terminal domain interface. It has been shown (61) that the isolated C-terminal domain of human γS-crystallin forms crystallographic dimers with a stable hydrophobic dimer interface composed of the same residues as the domain interface in the native state (**Fig. 5A**). We found that the HγD C-terminal domain also formed a dimer, eluting close to the full-length WT position by size-exclusion chromatography (**Fig. 5B**). Dimerization was confirmed by static light scattering in batch mode, as indicated. We propose that transient dimerization of HγD molecules in the open-interface intermediate state via formation of non-native intermolecular C-td:C-td interfaces eliminates stabilization of the N-terminal domain through interface with more stable C-terminal domain and therefore facilitates misfolding of the N-terminal domains. In the case of WT that opening of the interface is too rare and transient, and the N-terminal domain core too stable, to generate detectable misfolding of the N-terminal domain core to form the aggregation-prone 32-SS-41 non-native disulfide. However, when the N-terminal domain core is destabilized by the W42Q or other cataract-associated mutations – or when interface opening is a permanent feature, as in the case of the Δ85,86 variant – the relatively labile N-terminal domain misfolds to populate the aggregation-prone intermediate. In this way, interaction of two HγD molecules in similar intermediate states (open-interface) facilitates their conversion into a distinct and more unfolded intermediate state (hairpin-extruded) and thus accelerates aggregation. This structural model is summarized in **Fig. 6**.

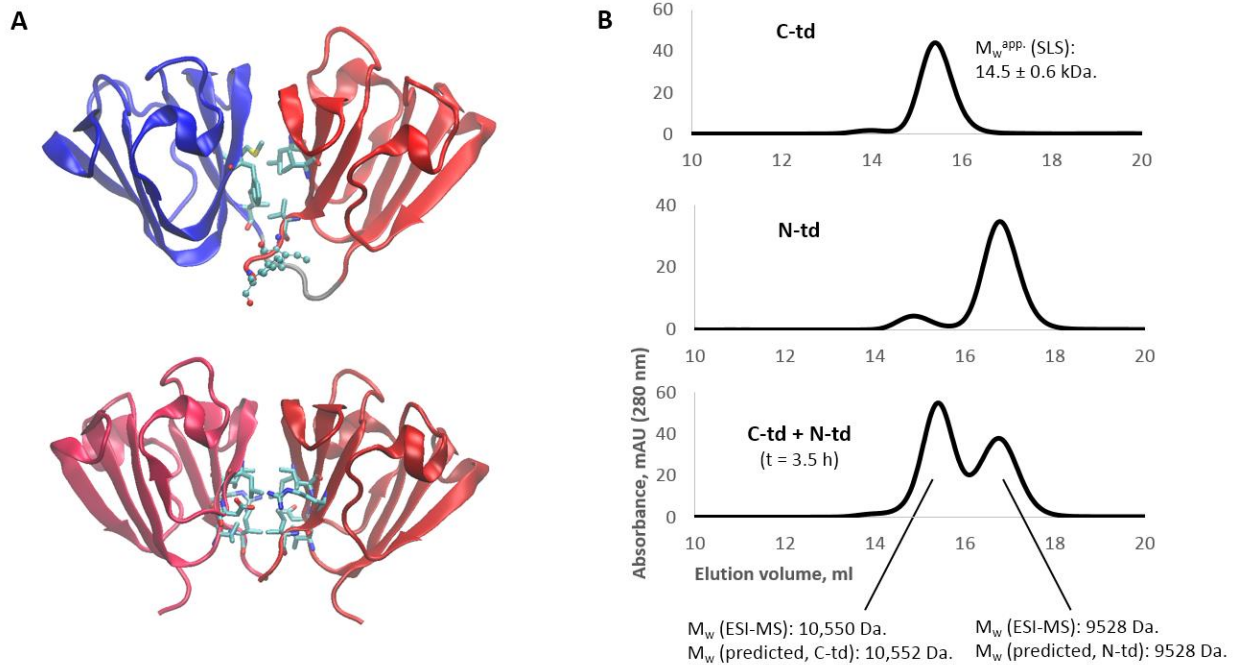

**Fig. 5: Homodimerization of the C-terminal domain is favored over N-td:C-td heterodimerization.** (A) A comparison between the native-state N-td:C-td domain interface of HγD (PDB ID 1HK0) (49) and the interface of the HγS C-td dimer (PDB ID 1HA4) (61). Ile171 and Phe172, in ball-and-stick representation in the HγD structure, would become available as part of a putative C-td-C-td interface. (B) Size-exclusion chromatography of isolated C-td (*top*), isolated N-td (*middle*), and a mixture of N-td and C-td following a 3.5 h incubation at 37 °C in the presence of DTT and ETDA. The mixture experiment did not result in an elution peak containing both N-td and C-td; rather, the two domains eluted separately, N-td as a monomer and C-td as a dimer, as confirmed by electrospray mass spectrometry. The C-td:C-td dimer therefore appears to outcompete the native-like N-td:C-td pairing.

## Discussion

In previous work we have shown that both misfolding and formation of a nonnative internal disulfide are required for the aggregation process in cataract-associated variants of HγD (41, 53). This disulfide locks a specific aggregation-prone structure, and formation of the same disulfide was found to correlate with cataract severity (33). We have also shown the surprising ability of HγD, and likely other γ-crystallins, to dynamically exchange disulfide bonds, potentially serving as a redox buffer (42), for which there is also supporting evidence from lens proteomics (35, 40). The "hot potato" mechanism proposed in (42) asserts that WT HγD has an oxidoreductase activity by oxidizing other HγD molecules in a relay-like manner until a mutationally or post-translationally destabilized HγD molecule gets oxidized with subsequent formation of the 32-SS-41 disulfide causing irreversible aggregation and elimination of the disulfide from solution. This way oxidized WT promotes aggregation of W42Q or other destabilized variants.

We have now shown that the WT remains a potent catalyst of W42Q aggregation even under strongly oxidizing condition (large excess of GSSG) suggesting an additional mechanism by which WT HγD can promote aggregation of mutant aggregation prone variants. Conversely, even oxidoreductase-inert versions of WT HγD (CCCC) dramatically increased aggregation kinetics and conformational dynamics of cataract-associated W42Q HγD. Aggregation of other cataract-associated variants (V75D and L5S) was accelerated even more. These findings suggest a novel mechanism by which WT promotes aggregation of cataract-associated variants – *conformational catalysis*. In the conformational catalysis mechanism, partial unfolding of one HγD molecule catalyzes misfolding of another. The observation of strong C-td:C-td association, combined with prior findings of significant destabilization of the isolated N-td, led us to propose that in HγD conformational catalysis proceeds by "interface stealing." Our kinetic model based on the assumption that conformational catalysis accelerates monomolecular conversion of the native state to the aggregation-prone intermediate in the cataract-associated variant predicted a robust (independent of kinetic parameters) functional form of dependence of aggregation kinetics on initial concentration of aggregating and catalyst variants, and the data fully supported this prediction (**Fig. 3**).

It's important to note the observed aggregation pathway does not depend on any particular mutation: oxidative aggregation under physiologically relevant conditions occurs not only for W42Q, but also for V75D, L5S, and even fully wild-type N-terminal domain, as long as the interface with the C-terminal domain has been disrupted. Disruption of the domain interface need not even require drastic damage, since it has been shown that one or two deamidations can be enough (51). These findings indicate that a variety of perturbations to diverse areas of the protein's structure may all converge on similar partially unfolded aggregation-prone intermediates.

Here we reported a peculiar structural mechanism of conformational catalysis which involves interface stealing by the catalyst protein from the aggregation-prone variant (**Fig. 5B**). Note that the two domains of the γ-crystallins arose by gene duplication and fusion (62); they are structurally identical, but their sequences have diverged (37% identity). Structural similarity promotes propensity of proteins to interact (63, 64). Further, it was shown that for structurally similar domains sequence divergence weakens their interactive propensity (65)consistent with our observation that homodimerization of C-td is more favorable than C-td:Ntd heterodimerization (**Fig. 5B**). The C-terminal domain of human γS-crystallin can form dimers (61), and in that protein, too, the resulting loss of the native domain interface greatly accelerates unfolding of the Cys-rich N-terminal domain (52, 60). Dimerization of the N-terminal domains via disulfides (66) may serve a protective function in this scenario. We expect that conformational catalysis like that presented here for HγD may also operate in HγS, other members of the γ-crystallin family, and perhaps other multidomain proteins.

In WT proteins interface stealing is likely not prevalent for entropic reasons – due to the high effective concentration of the two domains tethered by only a 5-residue linker. Furthermore, it has been shown that protein stability contributes to the strength of interaction with other proteins (67, 68). Thus, when the N-td is destabilized as in aggregation-prone variants, the stability of N-

td:C-td interface is likely also weakened, making the intermolecular C-td:Ctd interface relatively more favorable. This is especially true for mutations at Trp42, which disrupt both the N-td core and the domain interface (55). This double disruption explains why W42Q is much more aggregation-prone than L5S or V75D despite relatively small differences in thermostability (**Table 1**) – the C-td:C-td mode of dimerization is especially favorable in W42Q, making it a potent self-catalyst. The L5S and V75D mutations are not at the domain interface, so their rates of self-catalyzed aggregation are lower, and aggregation is accelerated by the WT protein to a relatively greater extent.

The phenomenon of conformational catalysis is conceptually the opposite of the by now well-known case of conformational templating (as in prions). In the templating case, protein-protein interactions shift one protein molecule into the same conformation as another (69-71). In the present case, partially unfolded WT protein catalyzes the mutant's conformational conversion to a state that the WT itself does not populate. "Inverse prion" interactions of this type have not been reported to our knowledge previously. Yet, they are likely to play a role in determining the properties of highly concentrated protein solutions. Conformational catalysis is expected to cause conformational heterogeneity to arise more rapidly at high protein concentrations than at low concentrations. This may promote non-native aggregation, but may also inhibit deleterious native-state aggregation, such as crystallization, which is an important evolutionary constraint in the lens (26, 49, 72). Future research should examine whether any defensive mechanisms have evolved to mitigate the aggregation-promoting effect of conformational catalysis.

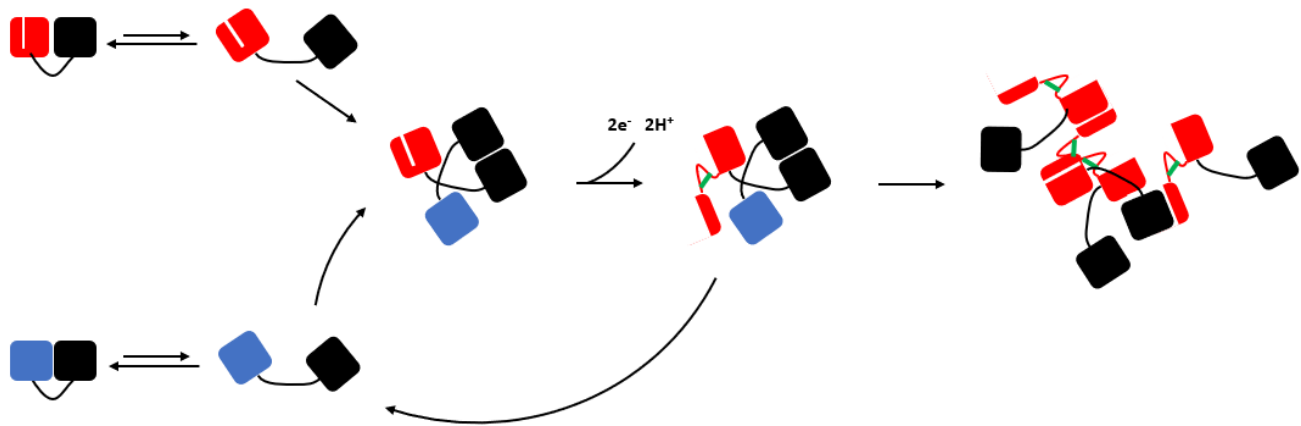

**Fig. 6: Structural model of conformational catalysis.** The catalytic interaction depends on a complex between cognate domains of two molecules (black), which deprives their partner domains (red, blue) of the added kinetic stability contributed by the native domain interface. If this partner domain also bears a destabilizing modification (white stripe), misfolding becomes likely. In the case of HγD, the misfolded conformer is locked by a non-native disulfide bond (green), while the wild-type intermediate is free to dissociate without further misfolding in a reaction driven by re-formation of the native domain interface. The aggregate model depicts the three recognized interaction modes between the detached N-terminal hairpin of HγD W42Q and sites on other copies of the protein, based on (41).

The general requirements for conformational catalysis to exist are relatively few. The mechanism presented here requires a multidomain protein where the stability of one domain is derived in part from its interface with another, and where interface stealing is possible. Besides interface stealing, classical domain swapping could also lead to a similar effect, as long as geometric or other constraints permit only a single swap, leaving one of the two destabilized domains unpaired. More generally, whenever a transient interaction between two protein molecules in a non-native conformational state further alters the conformation of one of them, conformational catalysis may be said to occur. Future research should investigate how significant a role conformational catalysis, and interface stealing in particular, may play in the stability, function, and aggregation of multidomain proteins.

## Methods

**Protein expression and purification** Expression and purification were carried out as described in (42). Briefly, untagged HγD constructs were expressed from the pET16b vector in BL-21 RIL cells in SuperBroth medium (Teknova). The cells were grown to late-log or early-stationary phase to obtain the highest yields under our expression conditions. After lysis by sonication, the proteins were purified by ammonium sulfate fractionation (collecting the fraction between 30% and 50% ammonium sulfate), followed by ion exchange on a Q-sepharose column (GE Life Science) and size exclusion on a Superdex75 26x600 column (GE Life Science) in the standard sample buffer, 10 mM ammonium acetate with 50 mM sodium chloride, pH 7. The three mutants whose aggregation traces were used in Figure 3 were treated with a reducing agent (1 mM dithiothreitol) for 2 h at room temperature prior to size exclusion.

**Solution turbidity assays and fitting** Turbidity assays were carried out as described in (42). All samples contained 1 mM EDTA to inhibit any trace metal-induced aggregation as seen in (73-75). Sample volume was 100 μl in half-area polypropylene plates (Greiner Bio-One), resulting in a path length of ~0.4 cm at the meniscus. Readings were taken in 96-well format on the PowerWave HT plate reader (BioTek). Maximum rate of turbidity was defined as the slope of the steepest tangent, which occurred early in the turbidity trace. Threshold values of turbidity were determined empirically for each mutant; linear regression fits were applied to the 10 time points starting 1, 2, and 3 points post-threshold. The resulting slopes were averaged. Fits to the maximum rate vs. concentration data used linear regression of the form $y = ax^2$, with $a$ as the only fitting parameter. Turbidity thresholds were set at 0.04 for W42Q; 0.02 for V75D; and 0.005 and 0.1 for L5S, since the aggregation of that variant appeared biphasic (see **Fig. S4**).

**PEGylation assays** Cys-accessibility PEGylations were carried out as follows. W42Q (20 μM) was incubated with or without the C18T/C78A/C108S/C110S variant (100 μM) and PEG(5000)-maleimide (500 μM) (Sigma) in standard sample buffer at 37 °C in the absence of any oxidizing agent. 5-μl samples were taken at the time points indicated (**Fig. 2** and **Fig. S3**) and mixed with 15 μl of denaturing buffer containing 5 μl 4x Tris/SDS NuPAGE gel-loading buffer (Thermo Fisher), 10 mM tricarboxyethyl phosphine (TCEP) as the reducing agent, and 1 mM L-Cysteine as the quencher of PEG-maleimide. The denaturing buffer was prepared the day of the experiment and allowed to equilibrate for 1 h prior to use to ensure all L-Cysteine was in its reduced form. Three or four replicate samples were used in each case. Cys-counting PEGylations were carried out as described in (42), in the presence of reducing agent (1 mM TCEP). Minor bands at +4PEG in Figure 1E indicate the reduction of internal disulfide bonds in W42Q was still not 100% complete, possibly because some bonds remained buried during the reductive treatment.

**Analytical size-exclusion chromatography** Samples at 10 μM concentration in standard sample buffer were incubated at 37 °C with 1 mM DTT and 1 mM EDTA for between 2 and 3.5 h. In the case of the N+C mixture the concentration was 10 μM of each. 100 μl of each sample was injected on the Superdex 75 10x300 analytical column (GE Life Science) in standard sample buffer, and peak fractions were collected manually.

**Differential scanning calorimetry** A nanoDSC robotic calorimeter (TA Instruments) was used, following the procedure in (41). Samples were buffer-exchanged by dialysis to 20 mM sodium phosphates, 50 mM NaCl buffer. Reducing agent (1 mM TCEP) was added prior to the experiments.

**Mass spectrometry** MALDI-TOF mass spectrometry was carried out following the protocol of (76) using a stainless steel target plate (Bruker) and the Bruker Ultraflextreme mass spectrometer. Pelleted samples were partially resolubilized with 0.1 M pH 5 ammonium acetate buffer. Samples were desalted using Pierce C18 tips (Thermo Fisher) and eluted in 80% acetonitrile with 0.1% trifluoroacetic acid. For electrospray mass spectrometry, samples from the size-exclusion were desalted using Pierce C18 tips (Thermo Fisher), eluted in 80% acetonitrile with 1% formic acid, and analyzed by LC/MS on an Agilent 6210 electrospray mass spectrometer, followed by maximum entropy deconvolution with 1 Da step and boundaries of 7,000-25,000 Da.

**Static light scattering** Samples at concentrations of 40 μM or (for C-td) 100 μM in standard sample buffer were treated with 1 mM DTT for ~1 h at room temperature to remove any possible C110-C110 dimers. They were then analyzed in 50-100 μl volume on the NanoStar dynamic light scattering instrument (Wyatt) in static light scattering mode, using disposable cuvettes from the manufacturer. The autocorrelation functions were deconvoluted using the Wyatt Dynamics software with the following restraints: 1.5-200 μs correlation times and 1.0-4.0 nm hydrodynamic radii. At least 20 measurements were averaged for each sample.

## Supplementary Information

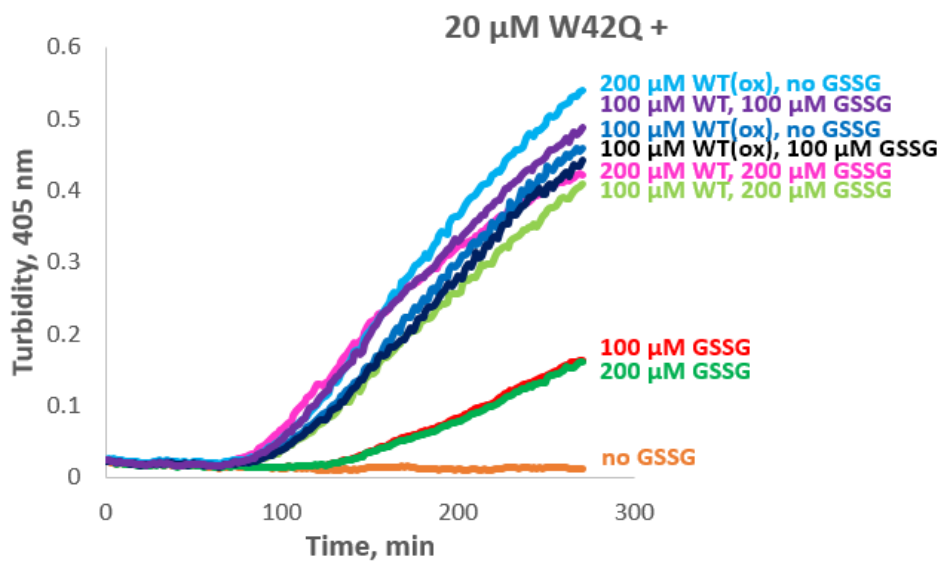

**Fig. S1: Oxidation status of WT does not affect its ability to catalyze W42Q aggregation when excess GSSG is present.** We have previously shown (ref. 42) that in the absence of any oxidation WT does not promote aggregation of W42Q, and that this aggregation requires non-native disulfide formation in W42Q (ref. 41). GSSG alone promoted W42Q aggregation, and so did oxidized WT, both consistent with these prior reports. However, the effect of GSSG + WT, or the WT protein alone if oxidized, was far greater than that of GSSG alone at the same concentration. Indeed, oxidation status of WT made no difference if excess GSSG was present in the sample.

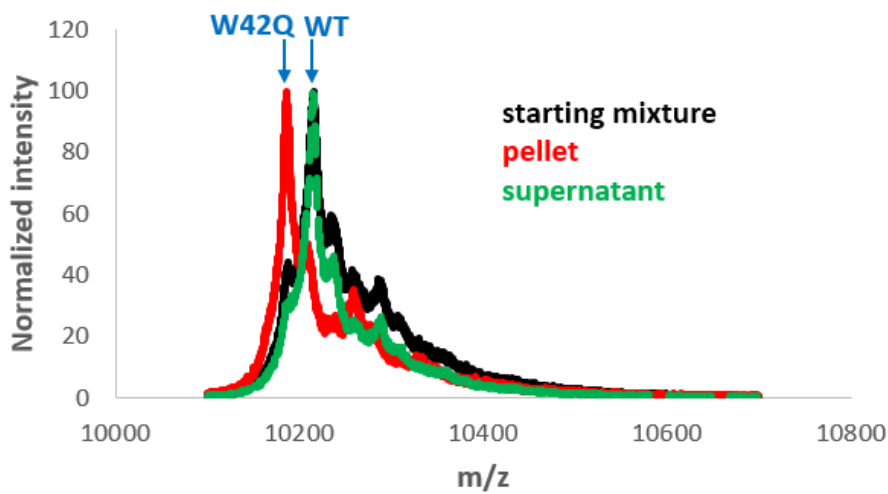

**Fig. S2: Aggregates formed in the WT + W42Q + GSSG mixture contain only W42Q.** MALDI-TOF spectral region containing the z = 2 ion confirmed that the starting mixture (*black*) contained both W42Q and WT, with an excess of the latter; the pelletable fraction contained only W42Q (*red*); and the remaining supernatant (*green*) contained the same amount of WT, but less W42Q, compared to the starting mixture.

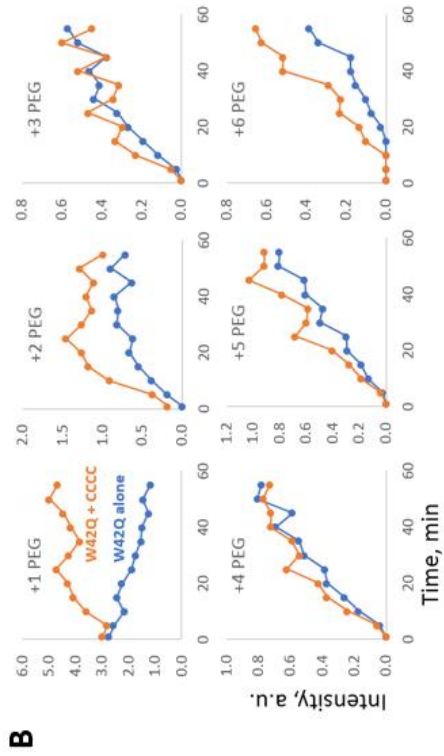
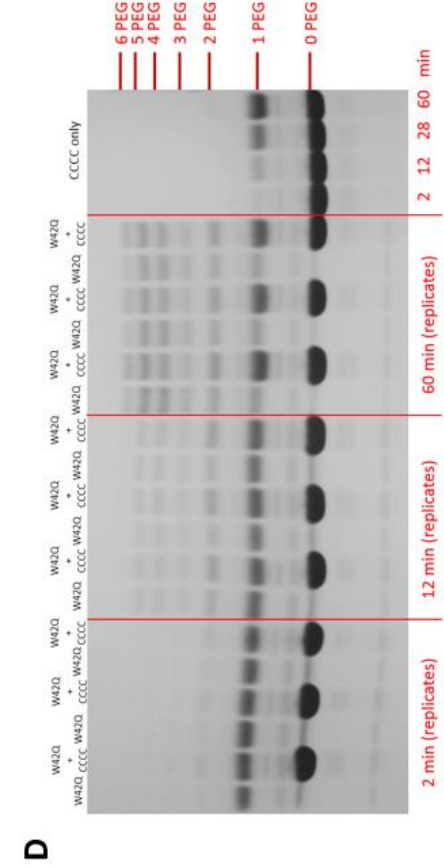
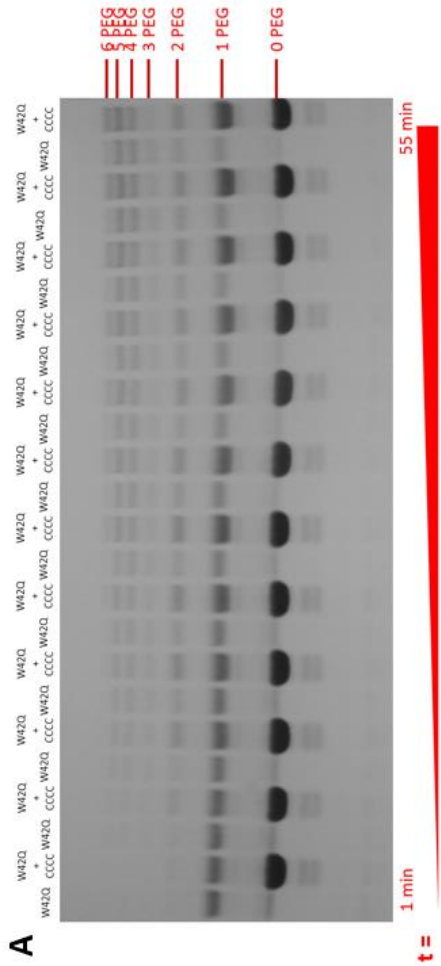
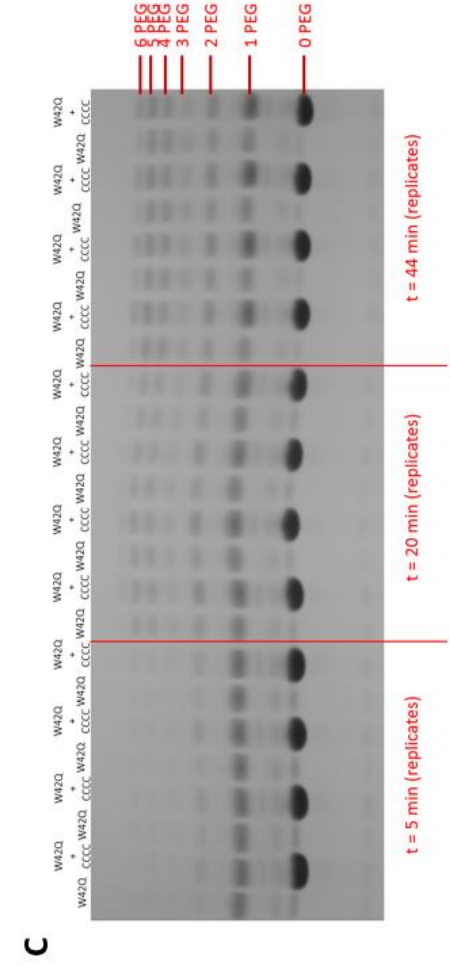

**Fig. S3: Solvent accessibility of buried Cys residues in the W42Q N-terminal domain increases in the presence of WT analog.** (A, C, D) The W42Q mutant sample was incubated with the WT-like quadruple Cys variant C18T/C78A/C108S/C110S and PEG-maleimide, and samples taken at indicated time points were analyzed by SDS-PAGE with subsequent quantitation by gel densitometry using GelAnalyzer 2010 software (B). Quantitation of the gels shown in (C) and (D) is presented in **Fig. 2**.

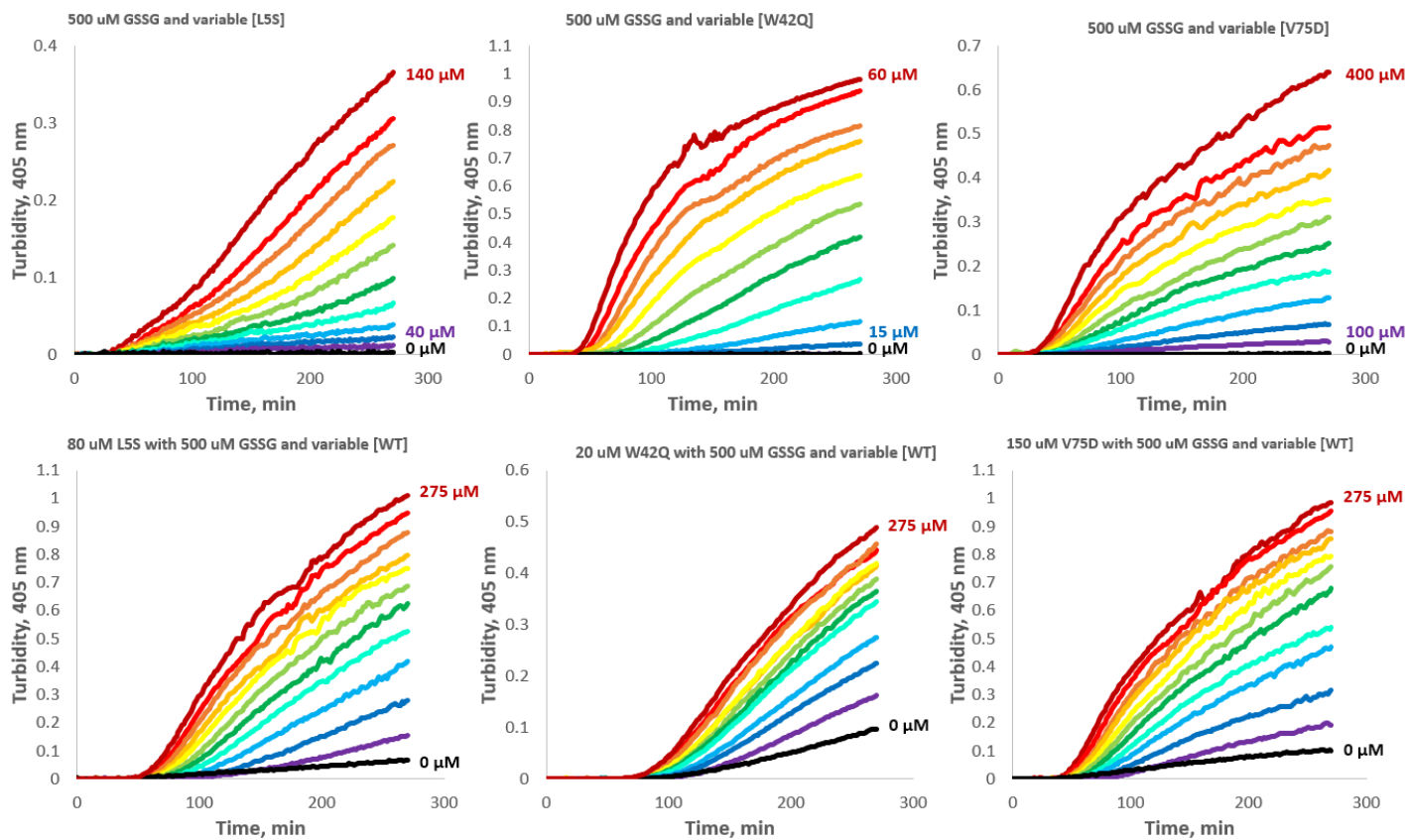

**Fig. S4: Turbidity traces for homogeneous and heterogeneous (WT-catalyzed) aggregation of the L5S, W42Q, and V75D variants.** Concentration of the variable protein increases from blue to red, with the lowest and highest concentrations indicated. *Top row*: Aggregation of L5S was assayed at concentrations from 40 to 140 µM, with a step of 10 µM; W42Q was assayed from 10 to 60 µM, with a step of 5 µM; V75D was assayed from 100 to 400 µM, with a step of 30 µM. *Bottom row*: Variable amounts of WT protein were added to a fixed amount of each mutant. [WT] was varied from 0 to 275 µM, with a step of 25 µM.

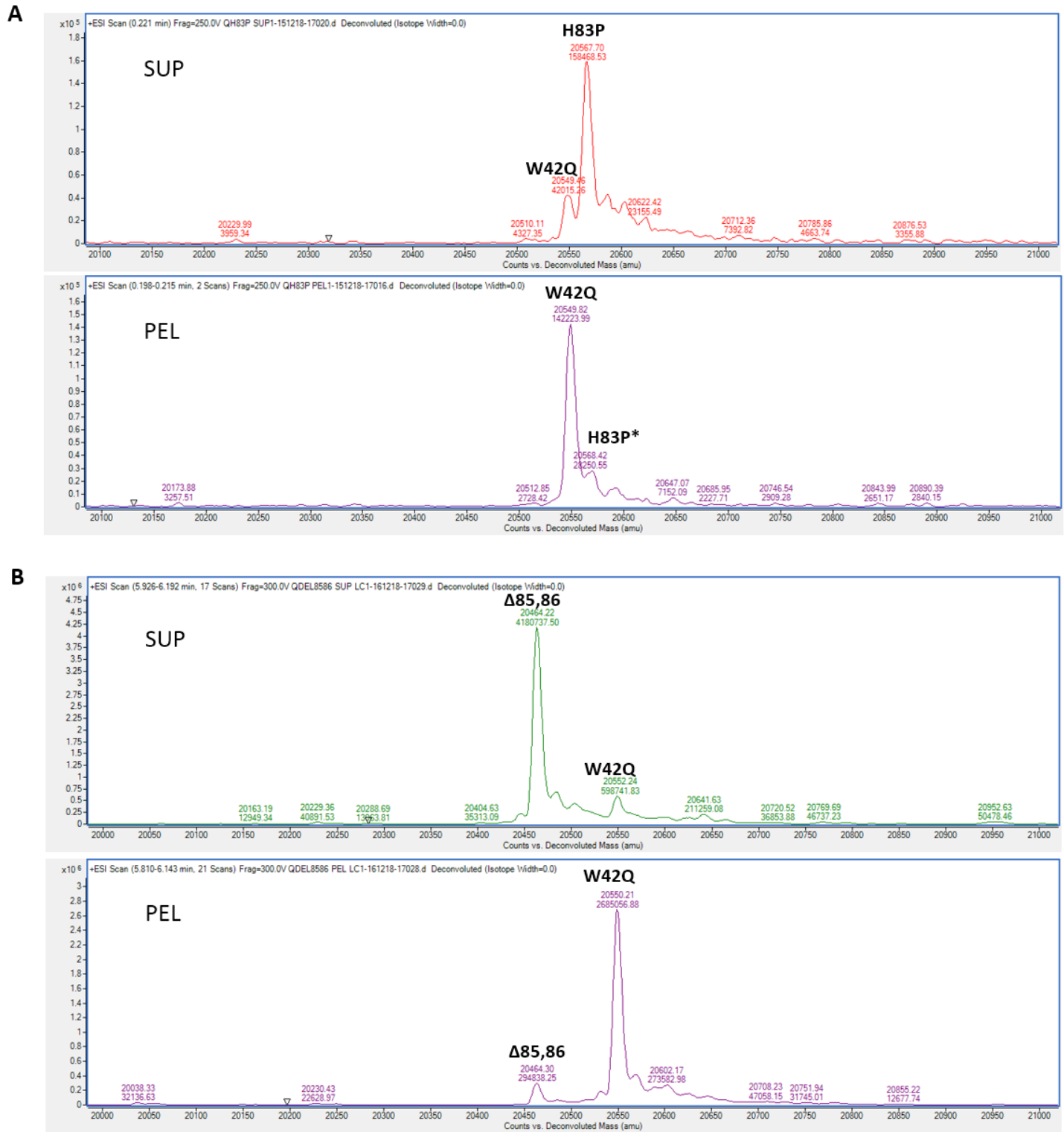

**Fig. S5: Coaggregation was observed between W42Q and Δ85,86** (*B*), **but not between W42Q and H83P** (*A*) in deconvoluted whole-protein electrospray mass spectra. Supernatant ("SUP") and pellet ("PEL") fractions from each aggregation experiment were analyzed. *Since their masses are too similar, the W42Q single oxidation (+16 Da.) peak and the native H83P peak cannot be reliably resolved on the Agilent 6220 instrument. However, comparison

to the PEL sample in (*B*), where no H83P is present, suggests that any co-aggregation of W42Q with H83P is at or below the limit of detection.

**Table S6: Solvent-accessible hydrophobic surface area of HγD WT compared to its isolated domains and to BSA**

| Protein | SASA, Å$^2$ |
|---|---|
| HγD WT | 474 |
| HγD N-terminal domain | 1126 |
| HγD C-terminal domain | 971 |
| BSA | 5415 |

Visual Molecular Dynamics (VMD) software (U. Illinois) was used to calculate the sum of solvent-accessible surface area belonging to W, F, L, I, V, M, A, or P residues in HγD WT (PDB ID 1hk0), its isolated domains (by removing the cognate domain from the same PDB structure), and bovine serum albumin (BSA) monomer (PDB ID 3v03).

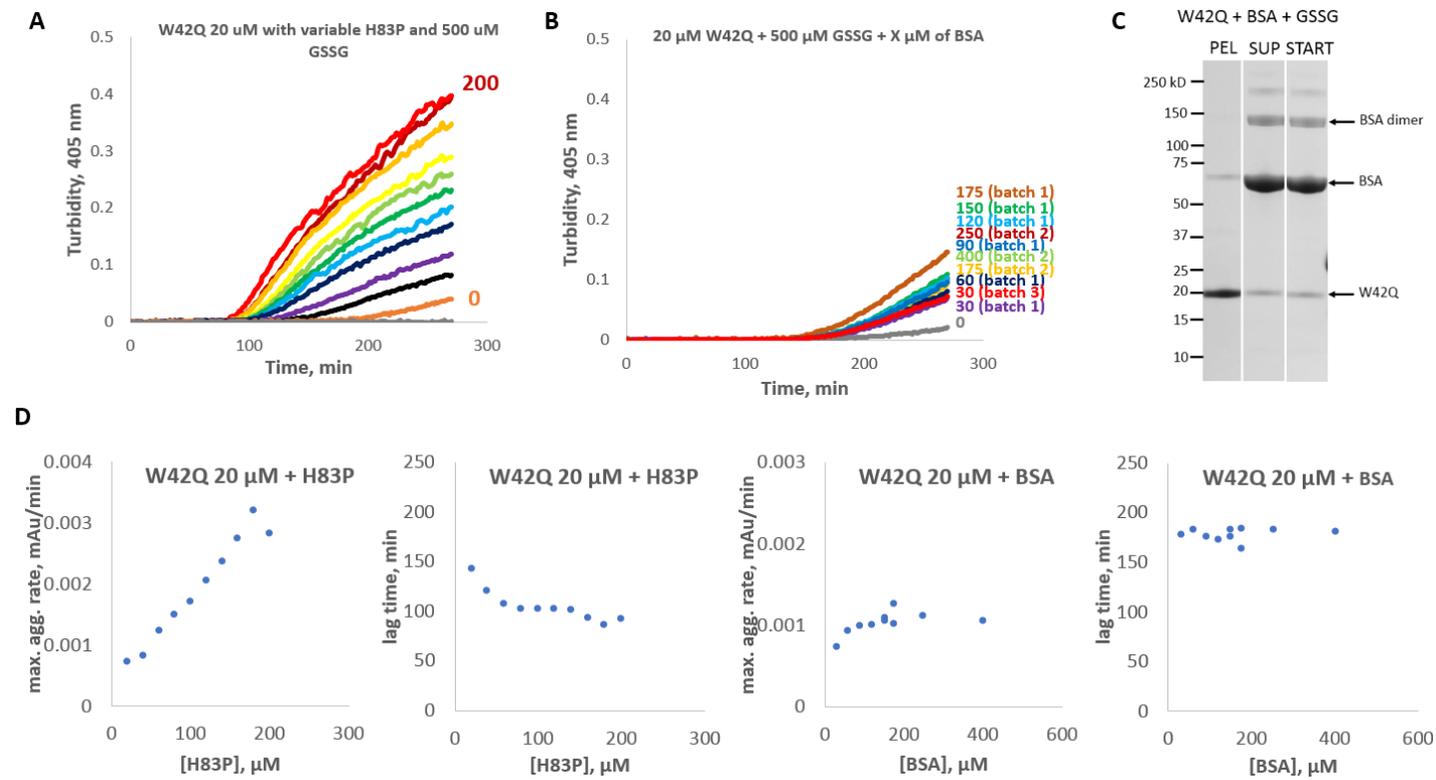

**Fig. S7: Conformational catalysis is not due to surface hydrophobicity.** (*A*) Aggregation traces of W42Q with variable concentration (from 0 to 200 µM, with a step of 20 µM) of H83P revealed much greater turbidity than (*B*) the same in the presence of variable [BSA] from three separate batches as indicated. The BSA protein was prepared each time by dissolving several milligrams of BSA powder (Sigma), followed by size-exclusion chromatography, isolation and concentration of the monomer fraction. (*C*) Modest co-aggregation was observed between BSA and W42Q by pellet/supernatant separation. This could be attributable to disulfide-mediated cross-linking between the two proteins. (*D*) While presence of BSA did accelerate the maximum aggregation rate of

W42Q, the effect was both smaller and much less dose-dependent than for H83P; furthermore, the lag time of aggregation, defined as the x-intercept of the steepest tangent, was significantly longer in the case of BSA.